\documentclass{ws-p8-50x6-00}

\begin{document}
 
\title{Higgs Searches and prospects from LEP2}

\author{Monica Pepe Altarelli}

\address{Laboratori Nazionali dell''INFN, Via E.Fermi 40, I-00044 FRASCATI\\
 and CERN, EP Division
\\E-mail: monica.pepe.altarelli@cern.ch}


\maketitle
\newcommand {\MW}      {M_{\mathrm{W}}}
\newcommand {\MZ}      {M_{\mathrm{Z}}}
\newcommand {\MH}      {M_{\mathrm{H }}}
\newcommand {\Mt}      {M_{\mathrm{t}}}
\newcommand {\Mh}      {M_{\mathrm{h}}}
\newcommand {\MA}      {M_{\mathrm{A}}}
\newcommand {\ff}      {{\rm f}\overline{\rm f}}
\newcommand {\bb}      {{\rm b}\overline{\rm b}}
\newcommand {\qq}      {{\rm q}\overline{\rm q}}
\newcommand {\elel}    {\ell^+\ell^-}
\newcommand {\tautau}  {\tau^+\tau^-}
\newcommand {\mumu}    {\mu^+\mu^-}
\newcommand {\nunu}    {{\nu}\overline{\nu}}
\newcommand {\bha}     {\mathrm{e}^+\mathrm{e}^-}
\newcommand {\WW}      {\mathrm{W}^+\mathrm{W}^-}
\newcommand {\roots}   {\sqrt{s}}
\newcommand {\Bh}      {\mathrm{B_h}}
\newcommand {\Mev}     {\mbox{$\rm MeV$} }
\newcommand {\Mevc}    {\mbox{$\rm MeV$}/c }
\newcommand {\Mevcc}   {\mbox{${\rm {MeV}}/c^2$} }
\newcommand {\Gev}     {\mbox{$\rm GeV$} }
\newcommand {\Gevc}    {\mbox{$\rm GeV$}/c }
\newcommand {\Gevcc}   {\mbox{${\rm {GeV}}/c^2$} }
\newcommand {\nb}      {\mbox{$\rm nb$} }
\newcommand {\pb}      {\mbox{$\rm pb$} }
\newcommand {\cm}      {\mbox{$\rm cm$} }

\abstracts{The status of the search at LEP2 for the Higgs in the 
standard model (SM) and in the minimal supersymmetric extension of
the standard model (MSSM) is reviewed. A preliminary lower limit of  
95.5~\Gevcc at 95\% C.L. on the SM Higgs is
obtained after a preliminary
analysis of the data collected at $\roots=189$~GeV.
For standard choices of MSSM parameter sets, the search for
the neutral Higgs bosons h and A leads to preliminary
95\% C.L. exclusion lower limits of 83.5~\Gevcc and 84.5~\Gevcc, respectively.
}

\section{Introduction}
After reviewing the indirect information on the Higgs mass based
on precise electroweak measurements performed at LEP1, SLD and at the 
TEVATRON, I will discuss the mechanisms of Higgs production
and decay and the strategy adopted to search for the neutral Higgs 
boson (in the SM and in the MSSM) at LEP2~\cite{reviews}.
I will summarise the results 
based on the analysis of approximately 
170~$\pb^{-1}$ collected by each LEP experiment at $\roots=189$~\Gev
updated to the more recent Winter Conferences numbers~\cite{felcini}.
In the end I will briefly discuss the prospects for Higgs
discovery at LEP2.

\section{Higgs mass from precision electroweak measurements and from theoretical arguments}
The aim of precision electroweak tests is to prove the
SM beyond the tree level plus pure QED and QCD corrections
and to derive constraints on its fundamental parameters.
Through loop corrections, the SM predictions for
the electroweak observables
depend on the top mass via terms of order $\rm{G_F}/\Mt^2$
and on the Higgs mass via logarithmic terms.
Therefore from a comparison of the theoretical predictions~\cite{pre_calc},
computed to a sufficient precision to match the experimental capabilities
and the data for the numerous observables which have been measured,
the consistency of the theory is checked and constraints on $\MH$
are placed, once the measurement of $\Mt$ from the TEVATRON is input. 
The present 95\%~C.L. upper limit on the Higgs mass in the SM is~\cite{mh_smfits,felcini}
\begin{equation}
\label{mh_up}
\MH< 220\,\Gevcc\,,
\end{equation}
if one makes due allowance for unknown higher loop uncertainties in the analysis.
The corresponding central value is still rather imprecise:
\begin{equation}
\MH= 71^{+75}_{-42}\pm5\,\Gevcc\,.
\end{equation}
The range given by Eq.\ref{mh_up} may
be compared with the one derived from theoretical arguments~\cite{hambye}.
It is well known  that in the SM with only one Higgs doublet a lower
limit on the Higgs mass $\MH$ can be derived from the requirement of vacuum
stability. This limit is a function of the energy scale $\Lambda$ where the 
model breaks down and new physics appears. Similarly an upper bound on 
$\MH$ is obtained from the requirement that up to the scale $\Lambda$ no 
Landau pole appears. 
If, for example, the SM has to remain valid up to the scale $\Lambda\simeq{\rm M_{GUT}}$,
then it is required that $135<\MH<180~\Gevcc$.

In the MSSM 
two Higgs doublets are introduced, in order to give masses to the 
up-type quarks on the one hand and to the
down-type quarks and charged leptons on the other.
The Higgs particle spectrum therefore consists of five physical states: 
two CP-even neutral scalars (h,A), one CP-odd neutral pseudo-scalar (A)
and a charged Higgs boson pair ($\rm{H}^{\pm}$).
Of these, h and A could be detectable at LEP2~\cite{yellow}. 
In fact, at tree-level h
is predicted to be lighter than the Z. However, radiative corrections
to $\Mh$~\cite{ellis}, which are proportional to the fourth power of the top mass,
shift the upper limit of $\Mh$ to approximately 135~\Gevcc, depending
on the MSSM parameters.

\section{Higgs production and decay}
At LEP2, the dominant mechanism for producing the standard model
Higgs boson is the so-called Higgs-strahlung process $\bha\to$~HZ~\cite{khoze,bjorken},
with smaller contributions from the WW and ZZ fusion processes leading to
H$\nu_{\rm{e}}\bar{\nu}_{\rm{e}}$ and H$\bha$ final states, respectively.
A sizeable cross section (few 0.1~pb) is obtained up to
$\MH \sim \roots - \MZ$, so that an energy larger than 190~\Gev is needed
to extend the search above $\MH \simeq \MZ$. 
For example the production cross section at $\roots=189$~GeV 
for $\MH=95$~\Gevcc is 0.18~pb, which
for an integrated luminosity $\cal{L}$=170~$\pb^{-1}$/exp. gives 30 signal events
per experiment.  

For the MSSM Higgs the main production mechanisms
are the Higgs-strahlung process $\bha\to$~hZ, 
as for the SM Higgs,
and the associated pair production $\bha\to$~hA~\cite{ha-prod}.
The corresponding cross sections may be written in terms of the
SM Higgs-strahlung cross section, $\sigma^{\rm{SM}}$, 
and of the cross section $\sigma^{\rm{SM}}_{\nunu}$ for the process $\rm{Z}^*\to\nunu$ as
\begin{eqnarray}
\label{Zh-hA}
\sigma(\bha\to\rm{Zh}) =  & \rm{sin}^2(\beta-\alpha)\,\sigma^{\rm{SM}} \\
\sigma(\bha\to\rm{hA}) \propto  & \rm{cos}^2(\beta-\alpha)\,\sigma^{\rm{SM}}_{\nunu}. \nonumber  
\end{eqnarray}
The parameter $\rm{tan}\beta$ gives the ratio of the vacuum expectation values
of the two Higgs doublets and $\alpha$ is a mixing angle in the CP-even sector.

The Higgs-strahlung hZ process occurs at large $\rm{sin}^2(\beta-\alpha)$, i.e., at
small $\rm{tan}\beta$. 
Conversely, at small $\rm{sin}^2(\beta-\alpha)$, i.e., at large $\rm{tan}\beta$,
when hZ production dies out, 
the associated hA production becomes the dominant mechanism with rates similar
to the previous case. In this region the masses of h and A are approximately equal.

For masses below $\sim 110~$~\Gevcc, the SM Higgs decays into $\bb$ in 
approximately 85\% of the cases and into $\tautau$ in approximately 8\%
of the cases. Similar branching ratios (BR) are expected for the MSSM Higgs bosons.
Above $\MH \sim 135$~\Gevcc, the BR into W and Z pairs becomes dominant.

\section{Searches at LEP2}
While at LEP1 energies 
the signal to noise ratio was as small as $10^{-6}$
due to the very high $\qq$ cross section, at LEP2 the signal to noise ratio
is much more favourable, increasing to $\simeq1\%$. In order to reduce this
background, mainly due to W pair production, $\qq$ (with two gluons or two
additional photons in the final state) and ZZ
events, use is made of b-tagging techniques which exploit
the large BR of the Higgs into $\bb$. For $\MH\simeq\MZ$, as is the case for
the expected experimental sensitivity, ZZ production represents
an irreducible source of background since the Z decays into $\bb$ in 15\% of the cases.

The following event topologies are studied: 
\begin{itemize}
\item[$ i)$] The leptonic channel (Z$\to \bha, \mumu$, H$\to\bb$) which represents
$7\%$ of the Higgs-strahlung cross section. These events are characterised by
two energetic leptons with an invariant mass close to $\MZ$ and a recoil mass equal
to $\MH$. Because of the clear experimental signature, no b-tag is necessary and therefore
the signal efficiency is high, typically $\sim75\%$.
\item[$ ii)$] The missing energy channel (Z$\to \nunu$, H$\to\bb$) comprising $\simeq20\%$ 
of the Higgs-strahlung cross section. This channel is characterised by a missing mass
consistent with $\MZ$ and two b-jets. The selection efficiency is $\simeq35\%$.
\item[$ iii)$] The four jet channel (Z$\to \qq$, H$\to\bb$)
which is not as distinctive as the two previous topologies but compensates for this 
drawback with its large BR of $\simeq64\%$. The efficiency for this channel is
typically $\simeq40\%$.
\item[$ iv)$] The $\tautau \qq$ channel (Z$\to \tautau$, H$\to\qq$ and vice-versa) with 
a $\simeq9\%$ BR. The event topology includes two hadronic jets and two oppositely-charged,
low multiplicity jets due to neutrinos from the $\tau$ decays. The signal efficiency is of the
order of 25\%. 
\end{itemize}
 
The b-tagging algorithms are based on 
the long lifetime of weakly decaying b-hadrons,
on jet shape variables such as charged
multiplicity or boosted sphericity and on high
$p_t$ leptons from semileptonic b decays.
The b-jet identification is  improved by combining
information from the different b-tagging algorithms with tools like
neural-networks and likelihoods. Typically,
for a 60\% signal efficiency, the WW background, which has no b-content,
is suppressed by a factor over 100, and the $\qq$ and ZZ backgrounds by
approximately a factor 10.
With respect to the b-tagging algorithms developed for the measurement at LEP1 of
$\rm{R_b}$, the b fraction of Z hadronic decays, 
the performances at LEP2 have
improved by almost a factor of 2, due to vertex detectors with an extended
solid angle coverage and to more efficient b-tagging techniques. 
 
All the analyses developed for the standard model Higgs 
produced via the Higgs-strahlung mechanism can be used with no modification
for the supersymmetric case, provided that the Higgs decays to standard model
particles ($\bb$, $\tautau$). The results can then be 
reinterpreted in the MSSM context, by simply rescaling 
the number of expected events by
the factor $\rm{sin}^2(\beta-\alpha)$.

For the pair production process, the signal consists of events with four
b-quark jets or a $\tautau$ pair recoiling against a pair of b-quark jets.

\section{Results and prospects}
Table~\ref{tab:res} shows the number of selected events in the data
for the SM Higgs search,
the expected number of background events and
the expected numbers of signal events
assuming $\MH=95$~\Gevcc~\cite{felcini,al_moriond,del_moriond,l3_moriond,op_moriond}.
\begin{table}[h]
\caption{Standard Model Higgs search. 
Number of observed events in the data $n_{\rm obs}$,
expected number of background events $n_{\rm back}$ and
expected numbers of signal events  $n_{\rm sig}$
assuming $\MH=95$~\Gevcc for the four LEP experiments 
and for their combination. Also shown are the number of events observed
and expected by the four experiments combined in the mass window
$\Delta\MH=92-96$~\Gevcc.}
\label{tab:res}
\begin{center}
\footnotesize
\begin{tabular}{|c|lll|}
\hline
& $n_{\rm obs}$     &$n_{\rm back}$ & $n_{\rm sig}$ \\
\hline
ALEPH        &  53 & 44.8 & 13.8 \\ 
DELPHI       &  26 & 31.3 & 10.1 \\
L3           &  30 & 30.3 &  9.9 \\
OPAL         &  50 & 43.9 & 12.6 \\
\hline
Total        & 159 & 150  & 46.4 \\
\hline
$\Delta\MH=92-96$~\Gevcc  & 47 & 37.5 & 24.6 \\
\hline
\end{tabular}
\end{center} 
\end{table}

As can be observed from Table~\ref{tab:res}, an excess of events 
is observed by ALEPH~\cite{al_moriond}
and OPAL~\cite{op_moriond} which, in the case of OPAL, is concentrated in the mass region around
$\MH\simeq\MZ$, while for ALEPH it is distributed over higher masses, typically $\geq95$~\Gevcc.
These results translate into the lower limits shown in Table~\ref{tab:lim}, together with the
sensitivity (expected limit) of each experiment. 
\begin{table}[t]
\caption{ Observed 95\% C.L. lower limits on $\MH$. Also shown are the limits predicted
by the simulation if no signal were present. 
}
\label{tab:lim}
\begin{center}
\footnotesize
\begin{tabular}{|c|cc|}
\hline
          & Observed & Expected \\
          & limit (\Gevcc)& limit(\Gevcc)\\
\hline
ALEPH     &    90.2 & 95.7 \\ 
DELPHI    &    95.2 & 94.8 \\
L3        &    95.2 & 94.4 \\
OPAL      &    91.0 & 94.9 \\
\hline
\end{tabular}
\end{center} 
\end{table}

%
%
%
Table~\ref{tab:MSSM_lim} shows the preliminary 95\% C.L. lower limits
on $\Mh$ and $\MA$ for the four LEP 
experiments~\cite{felcini,al_moriond,del_moriond,l3_moriond,op_moriond}, 
as well as the derived
excluded ranges of $\tan\beta$ for both  no mixing 
and maximal mixing in the scalar-top sector.
\begin{table}[b]
\caption{ Observed 95\% C.L. lower limits on $\Mh$ and $\MA$. Also shown are
the derived excluded ranges of $\tan\beta$. The mass limits are given for
$\tan\beta>1$, except for those of DELPHI, given for $\tan\beta>0.5$.
}
\label{tab:MSSM_lim}
\begin{center}
\footnotesize
\begin{tabular}{|c|cccc|}
\hline
          & $\Mh$ (\Gevcc) & $\MA$ (\Gevcc) & $\tan\beta$ & $\tan\beta$ \\
          &                &                & max. mixing  &  no mixing \\
\hline
ALEPH     &    80.8 & 81.2 & -                    & $1<\tan\beta<2.2$ \\ 
DELPHI    &    83.5 & 84.5 &  $0.9<\tan\beta<1.5$ & $0.6<\tan\beta<2.6$ \\
L3        &    77.0 & 78.0 &  $1.<\tan\beta<1.5$ & $1.<\tan\beta<2.6$ \\
OPAL      &    74.8 & 76.5 &  -                  & $0.81<\tan\beta<2.19$ \\
\hline
\end{tabular}
\end{center} 
\end{table}

In the years 1999 to 2000 LEP2 is expected to deliver
a luminosity larger than 200~$\rm{pb}^{-1}$ per experiment at 
a centre-of-mass energy eventually as high as $\sim 200$~GeV.
These data 
should allow to discover a SM Higgs of 107~\Gevcc or to exclude a Higgs lighter than
$\sim$108~\Gevcc~\cite{lellouch,chamonix}. 
This is a particularly interesting region to explore, given the present indication
for a light Higgs from the standard model fit of the electroweak precision data.
The sensitivity to the Higgs in the MSSM  will reach
$\sim90$~\Gevcc  for the high $\tan\beta$ region and $\sim108$~\Gevcc
for $\tan\beta\simeq1$, therefore allowing good coverage of the MSSM plane.
%

\section*{Acknowledgments}
I would like to thank  Cesareo Dominguez and Raul Viollier for their great
hospitality and excellent organization of the Workshop and
``Maestro'' Patrick Janot for his precious advice 
and for carefully reading this manuscript.

\end{document}